\documentclass[preprint,prb,aps]{revtex4}
\usepackage{graphicx, amssymb}
\begin{document}
\title{Solvent control of crack dynamics in a reversible hydrogel}
\author{Tristan Baumberger}
\email[corresponding author: ] {tristan.baumberger@insp.jussieu.fr}
\author{Christiane Caroli}
\author{David Martina}
\affiliation{INSP, Universit\'e Pierre et Marie Curie-Paris 6, 
Universit\'e Denis Diderot-Paris 7, CNRS, UMR 7588
Campus Boucicaut, 140 rue de Lourmel, 75015 Paris, France.}

\date{\today}
\maketitle

{\bf The resistance to fracture  of reversible biopolymer hydrogels is an 
important control factor of the cutting/slicing and eating 
characteristics of food gels\cite{Food}. It is also  critical 
for their utilization in tissue engineering, for which 
mechanical protection of encapsulated components is 
needed\cite{Tissue, Mooney}. Its dependence on loading 
rate\cite{McEvoy} and, recently, on the density and strength of 
cross-links\cite{Mooney} has been investigated. But no attention was 
paid so far to  solvent nor to environment effects. Here we report a 
systematic study of crack dynamics in gels of gelatin in 
water/glycerol mixtures. We show on this model system  that: (i) 
increasing solvent viscosity slows down cracks; (ii) soaking with 
solvent increases markedly gel fragility; (iii) tuning the viscosity of 
the (miscible) environmental liquid affects crack propagation via 
diffusive invasion of the crack tip vicinity. The results point 
toward the fact that fracture occurs by viscoplastic chain pull-out. 
This mechanism, as well as the related phenomenology,  should be common 
to all  
reversibly cross-linked (physical) gels.}

Gelatin gels are constituted of denatured (coil) collagen chains, held 
together by cross-links made of segments of three-stranded helices 
stabilized by hydrogen bonds\cite{Nij}. 
This network, swollen by the aqueous solvent, which controls its 
(undrained) bulk modulus, is responsible for the finite shear 
modulus $\mu$, of order a few kPa. Hence, hydrogels can be considered 
incompressible. One estimates average 
mesh sizes $ \xi  \sim (kT/\mu)^{1/3}$ of order $10$ nm, i.e. coil 
segments involving a few 100 units (residues)\cite{Courty}. 
Moreover, 
in the presence of pressure gradients, the solvent diffuses through 
the network. This poroelastic behaviour\cite{Onuki,DLJohnson} controls 
e.g. slow solvent draining in or out of the gel under applied 
stresses.

They are {\it thermoreversible}, i.e., in contrast with chemical, 
covalently cross-linked gels, their network "melts" close above room 
temperature. This behavior, assignable to their small cross-link binding 
energy, leads to the well studied\cite{Nij} slow aging (strengthening) of $\mu$, 
and to their noticeable creep under moderate stresses\cite{Higgs}. 
When stretched at constant strain rate, gelatin gels 
ultimately fail at a strain $\sim 1$ which, though rather poorly 
reproducible, is clearly rate-dependent\cite{McEvoy}. In order to get insight into 
the nature of the dissipative processes at play, one needs to 
investigate the propagation of cracks independently from their 
(stochastic) nucleation\cite{Bonn}. Here we study the fracture energy $\mathcal G(V)$ needed to propagate 
a crack at constant velocity $V$ in notched long thin plates (see 
Fig.~\ref{fig:Trace}) of gels differing by the glycerol content of their aqueous 
solvent.

As seen on Figure~\ref{fig:G(etaV)}(a), $\mathcal G$ is very strongly velocity 
and solvent-dependent. Our gels  are velocity toughening: at fixed 
$\phi$, $\mathcal G$ grows quasi-linearly over the whole investigated 
$V$-range (0.1--30 mm.s$^{-1}$).
Linear extrapolation down to $V = 0$ yields an evaluated 
quasi-static toughness $\mathcal G_{0}$. Within experimental 
accuracy, $\mathcal G_{0}$ is of order  $2.5$ J.m$^{-2}$, i.e. 
about $20$ times larger than a gel-air surface energy, and 
$\phi$-independent. In contrast, $\mathcal G(V)$ becomes 
noticeably steeper as $\phi$ increases. Plotting it versus 
$\eta V$ (Fig.~\ref{fig:G(etaV)}(b)) captures most of this 
dependence. 
Note that the ratio $\mathcal G/\eta V$ is a huge {\it 
number} of order $10^{6}$.

This points toward the critical role of network/solvent relative 
motion. In particular, the impossibility for not very thin,  quasi 
incompressible plates 
to accomodate fully the high local strain gradients developing in the 
crack tip region results in high negative pressures. So, most likely, 
solvent is partly drained out  of this region into the bulk, leading 
for the chains thus exposed to air, to a solvation energy cost. We investigate this issue with the 
help of experiments in which a drop of the gel solvent is introduced into 
the already moving crack opening. Such tip wetting induces, at fixed 
sample stretching, a positive, $V$-independent velocity jump 
(Fig.~\ref{fig:Gdry/wet}). 
Equivalently, wetting decreases ${\mathcal G}(V)$ by a constant 
$\Delta {\mathcal G}_{0}$. For $\phi = 0$, $\Delta {\mathcal G}_{0} 
\sim 2$ J.m$^{-2}$ is a substantial fraction of ${\mathcal G}_{0}$.

Can one make, on the basis of these results, a plausible guess about 
the nature of the fracture mechanism in reversible hydrogels?

Clearly, one cannot invoke here the classical Lake-Thomas picture 
\cite{LT}, which successfully accounts for rubber toughness. Indeed, 
in these materials, fracture occurs via chain scission:  polymer 
segments crossing the fracture plane are stretched taut until they 
store an elastic energy per monomer of order the covalent binding 
one, $U_{chain} \sim$ a few eV. In thermoreversible gels, the 
corresponding force , $\sim U_{chain}/a$, is more than two orders of 
magnitude larger than that, $f^{*} \sim U_{CL}/a$, which can be 
sustained by the $H$-bond stabilized cross-links (CL). $U_{CL}$ is 
the segmental unbinding energy  introduced in the zipper\cite{Nishi} 
and reel-chain\cite{Ball} models of gel elasticity. This leads us 
to postulate that, in the highly stressed "active crack tip zone", 
cross-links yield, up till the stretched chains are pulled out of the matrix. 
The threshold stress $\sigma^{*}$ at 
the onset of CL-yield can be estimated as $\sigma^{*} \sim f^{*}/\xi^{2} = 
U_{CL}/a\xi^{2}$. With $a \sim 0.3$~nm, $U_{CL} \sim 10^{-1}$~eV, 
$\xi \sim 10$ nm, 
$\sigma^{*} \sim 500$~kPa, two orders of magnitude larger 
than a small strain shear modulus.

When solvent can be pumped from a wetting drop, the plastic zone deforms under this 
constant stress till the opening $\delta_{c}$ at the tip reaches the 
length $\ell$ of a fully stretched chain. This Dugdale-like 
picture\cite{Lawn} yields for the quasi-static fracture energy :
\begin{equation}
\label{eq:dugdale}
{\mathcal G}_{0}^{wet} = \sigma^{*} \ell
\end{equation}
from which we get $\ell \sim 1.2\,\mu$m. 
With an average mass $M_{res} = 80$ Da per residue, this yields for the gelatin molar 
weight the reasonable estimate:  $M_{res}\ell /a\sim 320$ kDa.

Let us turn to the $V$-dependence of $\mathcal G$. A finite $V$ means 
a finite pull-out velocity 
$\dot\delta = \alpha V$. Determining the value and precise space 
variation of $\alpha = d\delta/dx$ in the active zone would demand 
solving for the whole stress field. Given the large value of 
$\sigma^{*}/\mu$, this would raise such intricate, still unsolved issues 
as strain-induced helix/coil transition\cite{Courty}, 
elastic crack blunting and strain hardening at large deformation 
\cite{Hui}.  
Failing anything better, we take $\alpha$ to be an 
unknown geometrical factor. 

We then write for the viscoplastic stress $\sigma = \sigma^{*} + 
\sigma_{vis}(V)$. The viscous stress $\sigma_{vis}$ 
can be evaluated as resulting 
from  
hydrodynamic friction of chains of contour length ${\ell}$ pulled 
at velocity $\dot{\delta}$ out of ``tubes" formed by the 
embedding network.
A natural candidate for the radius of these tubes is the effective 
pore size $\xi_{hydr}$ extracted from light scattering 
experiments\cite{Tanaka} (see Methods).  
So, 
$\sigma_{vis} \approx \alpha\eta V{\ell}/\xi_{hydr}^{2}$, and 

\begin{equation}
\label{eq:pente}
{\mathcal G}(V) \approx {\mathcal G}_{0} + {\ell}\sigma_{vis} = {\mathcal 
G}_{0} +\alpha\left(\frac{\ell}{\xi_{hydr}}\right)^{2}\eta V
\end{equation}

So, this schematic model does account for the linear ${\mathcal G}(V)$ 
variation. Moreover, with $\xi_{hydr} = 2.7$ nm, it predicts a reduced slope 
$d{\mathcal G}/d(\eta V) \approx 2.5\times 10^{5} \alpha$. When compared 
with experimental values ($\sim 10^{6}$) this suggests $\alpha$ 
values of order unity, possibly associated with elastic 
blunting\cite{Hui}.  

Beyond this, the remaining splay between the ${\mathcal G}(\eta V)$ 
curves (Fig.\,\ref{fig:G(etaV)}(b)) appears positively correlated 
with elastic stiffness variations: as is the case for reversible 
alginate gels\cite{Mooney}, ``the stiffer, the tougher" --- a 
relationship currenly under more systematic investigation.

In this picture, draining at a non-wetted tip results in a capillary 
energy cost  par chain $\Delta{\mathcal G}_{0}.\xi^{2} \sim 1000$ 
eV/chain, i.e. $\sim 10\,kT$ per residue, a value which suggests that 
chains are extracted individually rather than as gel fibrils. The 
observed $V$-independence of $({\mathcal G}^{dry} - {\mathcal 
G}^{wet})$ indicates that $\alpha$, hence the geometry of the active 
zone, is largely unaffected by wetting.

We can now take one further step. Our scenario suggests that we should 
be able to tune, at fixed grips, the crack velocity by using a wetting 
drop with a glycerol content $\phi_{drop}$, hence a viscosity, 
different from that of the gel solvent. We expect that, for small enough 
$V$, the miscible wetting fluid will
invade the whole active zone, bringing $V$ to 
its value for a $\phi_{drop}$-gel. The faster the crack, the less 
efficient this diffusive mixing process: the hetero-wetted ${\mathcal 
G}(V)$ should gradually approach that for the homo-wetted $\phi$-gel. 
The result for cracks in a $\phi = 30 \% $ gel wetted by pure water, 
shown on Figure 4, spectacularly confirms this expectation.

Moreover, assuming that the cross-over range $(V_{1}, V_{2})$ 
(Fig.~\ref{fig:HeteroWetting}) 
corresponds to diffusion lengths $D_{eff}/V$ decreasing from $d_{act}$ 
to $\xi$, we estimate for the diffusion constant of glycerol 
in the stretched gel $D_{eff} \sim V_{2}\xi \sim 2\times 10^{-10}$ 
m$^{2}$/s, and for the size of the active zone in the gel matrix 
$d_{act} \sim \xi V_{2}/V_{1} \sim 100$ nm.

We conclude from this work that due to their weak binding responsible 
for cross-link plasticity\cite{Higgs, Ball, Mooney}, thermoreversible gels fracture via 
chain pull-out. While the fracture threshold is controlled by the cross-link 
yield stress, i.e. by the network only, crack dynamics is ruled by 
chain/solvent friction. This opens promising perspectives toward 
solvent control of crack dynamics in these materials, since: (i)
The larger bulk solvent viscosity is, the slower cracks under a 
    given loading,  (ii) Homowetting of a crack tip speeds it up and can set subcritical precracks into 
    motion,  (iii) Heterowetting by a miscible fluid with substantial 
    viscosity contrast leads, via diffusive ``rinsing" of the active zone, to drastic 
    effects on the propagation of slow cracks.

So, tip wetting appears as a method of local and fast control 
    of crack dynamics in such materials. Conversely, our results call 
    attention to the fact that characterization of gel fracture 
    properties should be performed under realistic, e.g. 
    physiological, environmental conditions. 

\appendix

\section*{Methods}
\subsection*{Gel sample preparation}
Gels are prepared by dissolving 5 wt\% gelatin powder (type A from 
porcine skin, Sigma) in  mixtures of glycerol 
 ($\phi$ = 0, 20, 30, 60 wt\%)
in deionized water, 
under  
continuous stirring at 90$^\circ$ C, an unusually high temperature 
needed to get homogeneous pre-gel solutions at high $\phi$\cite{Ferry}. 
A control experiment performed with a pure water/gelatin sample prepared 
at 60$^\circ$ C resulted in differences on low strain moduli and   $\mathcal 
G (V)$ slopes of, respectively,   1\% and  7\%, compatible with 
scatters between samples prepared at 90$^\circ$ C, whence we conclude that our preparation method 
does not induce significant gelatin hydrolysis.

The solution is then poured into a mold 
consisting of a rectangular frame and two plates covered with Mylar$^{\scriptsize\circledR}$ films.
On the longest sides of the frame, the curly half-part of an adhesive 
Velcro$^{\scriptsize\circledR}$ tape improves the gel plate grip.
The mold is set at $2\pm 0.5^\circ$ C for 15 hrs, then clamped to the mechanical testing set-up 
and left at room temperature 
(19 $\pm 1^\circ$ C) 
for 1 hr. 
The removable pieces of the mold are subsequently taken off, leaving the $300\times 
30\times 10$ mm$^{3}$
gel plate fixed to its grips. The Mylar$^{\scriptsize\circledR}$ films are left in position in 
order to prevent solvent evaporation. They are peeled off just before 
performing an experiment.  

\subsection*{Hydrogel characterization}

Small strain shear  moduli $\mu$ are computed from 
the force-elongation curves assuming incompressibility, plane stress 
deformation and 
neglecting finite-size effects. A high 
glycerol content significantly stiffens the gels , from $\mu$ = 3.5 
kPa at $\phi$ = 0 to $\mu$ = 5.2 
kPa at $\phi$ = 60 \%, possibly due to $\phi$-dependent solvent-chain 
interactions.
Large deformation, non-linear curves up to stretching ratios $\lambda \simeq 
1.5$ are integrated numerically to determine the  elastic energy $\mathcal F(\lambda)$ stored 
in stretched plates.   

The collective diffusive mode of the 
polymer network in the solvent\cite{Onuki, DLJohnson} is charaterized by the diffusion 
coefficient $D_{coll}$, measured by dynamic light 
scattering as described elsewhere\cite{Gelpulses}.  
Solvent viscosities $\eta$ range from $10^{-3}$ Pa.s at 
$\phi$ = 0 to $11\times10^{-3}$ Pa.s  
at $\phi$ = 60 \%. Accordingly, $D_{coll}$ decreases as $\phi$ increases.
One estimates an effective pore size as 
$\xi_{hydr}= \sqrt{D_{coll}\eta/\mu}$. It is found independent of 
$\phi$ within experimental accuracy~: 
$\xi_{hydr} = 2.7 \pm 0.2$ nm.

\subsection* {Fracture experiments}

\begin{figure}[tbp]
    \centering
    \includegraphics[scale=0.5]{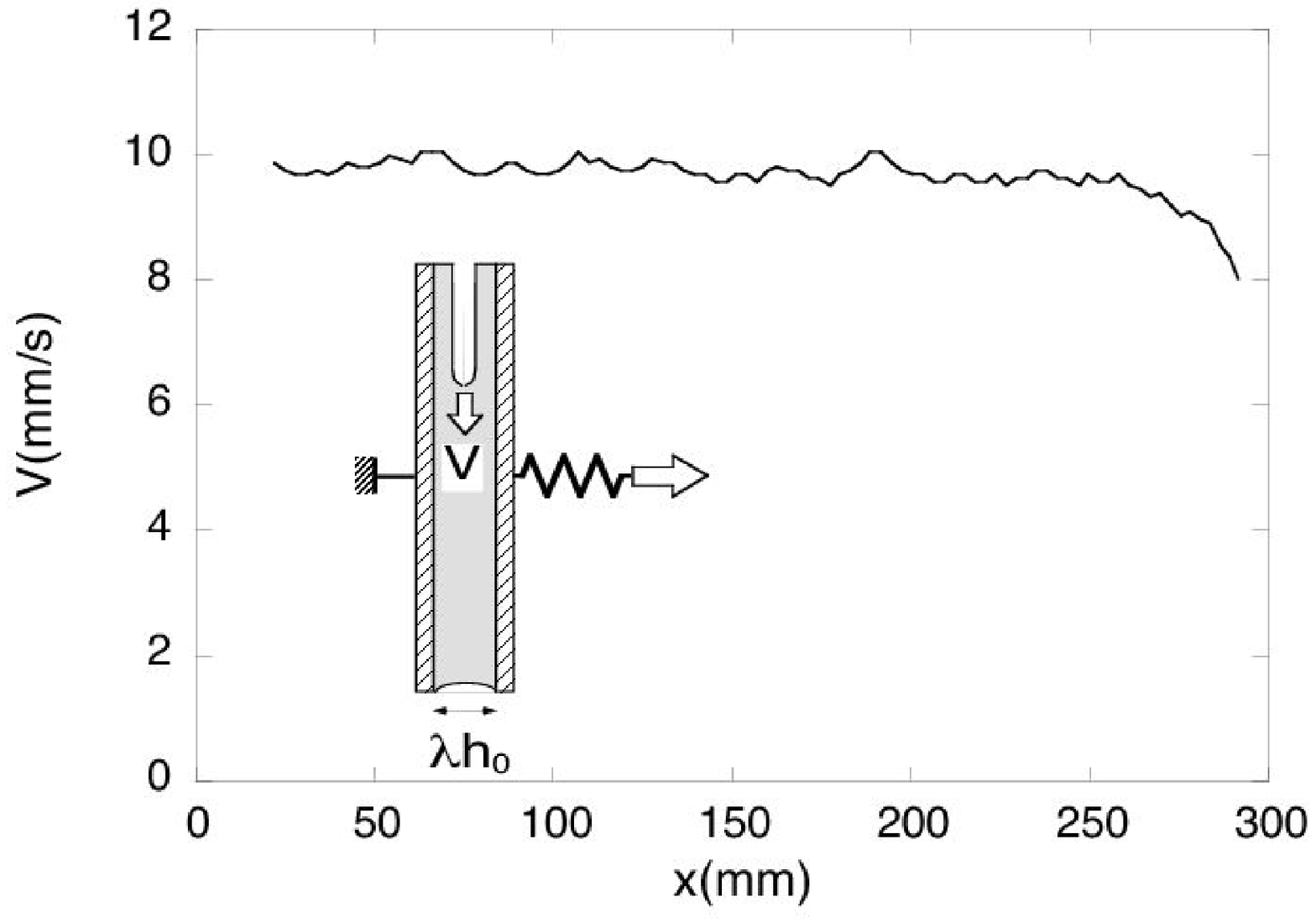}
    \caption{{\bf Velocity of a stable crack propagating in the mid-plane 
    of a long plate}.
    Sample: 5 wt \% gelatin gel in water, 10 mm thick, 300 mm  long, stretched in the transverse direction 
    by a factor $\lambda = 1.3$. The crack was initiated by a single 
    knife-cut notch, 20 mm long.  Edge effects extend over about $2h_{0}$ 
    with $h_{0}$ = 30 mm the width of the unstretched sample. The 
    remaining central part behaves as a homogeneously stretched plate, hence the 
    observed steady crack-propagation.}
    \label{fig:Trace}
\end{figure}

Before stretching, a knife-cut notch is made at one edge of the plate. 
The grips are then pulled apart for 1 second by an 
amount $\Delta h = \lambda h_{0}$, with $h_{0} = 30$ mm the height of the 
plate. The stiffness of the load cell is such that  fracture occurs at 
fixed grips.     
A video movie of the plate is recorded at  a typical 15 frame.s$^{-1}$ 
rate and post-treated for tracking the crack tip position. 
Away from the sample edges, cracks run at a constant velocity $V$ (see 
Fig.\,\ref{fig:Trace}).  
Further data processing is 
restricted to this region. $V$ is computed by linear regression of the tip position. 
Since reversible gels creep, the investigated  range is  
restricted to velocities large enough for the energy released by stress relaxation of the 
gel to be negligible compared with that released by crack propagation. 

\begin{figure*}[tbp]
    \centering
    \includegraphics[scale=0.5]{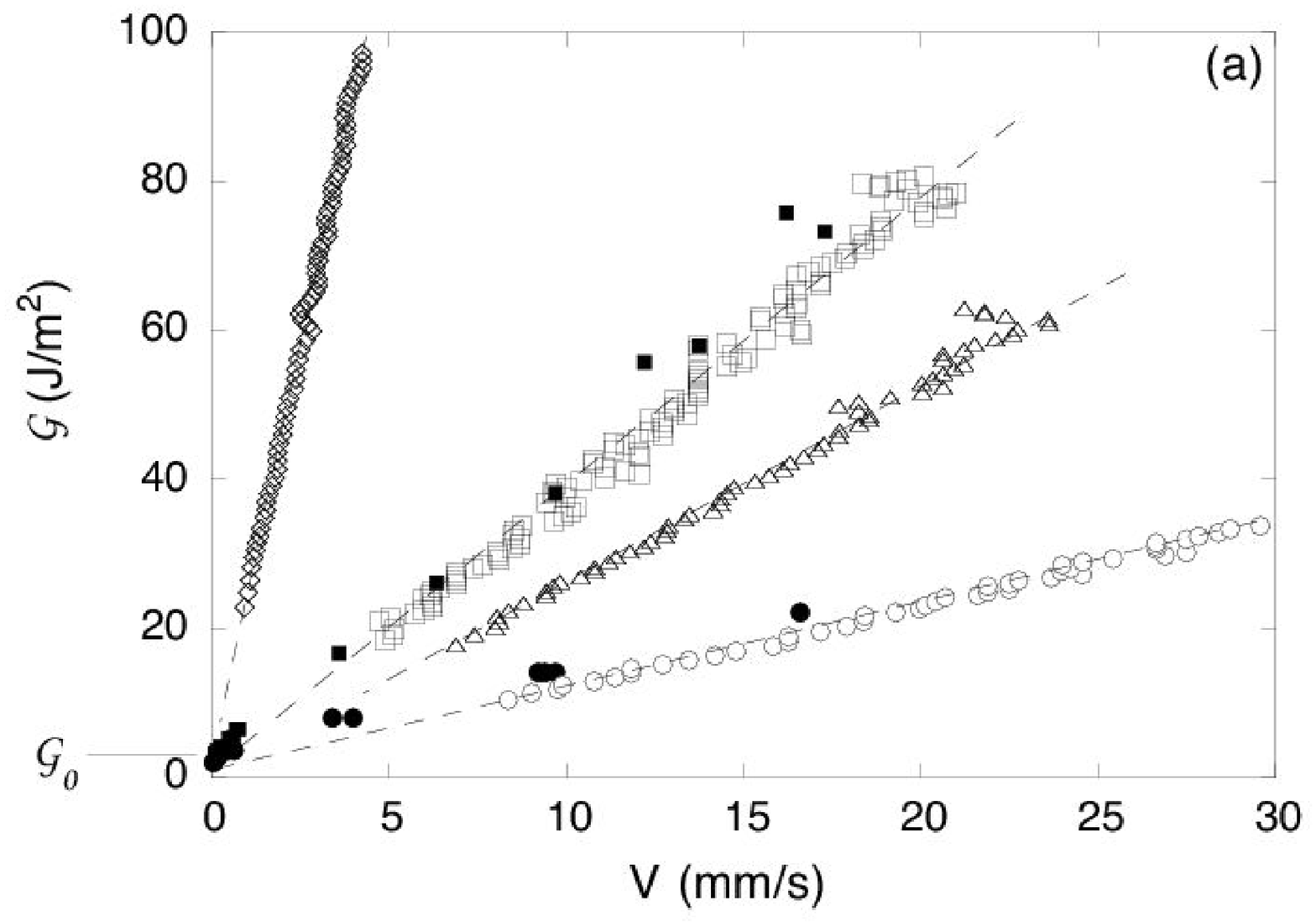}
      \label{fig:G(V)}
\end{figure*}

\begin{figure}[tbp]
   \centering
    \includegraphics[scale=0.5]{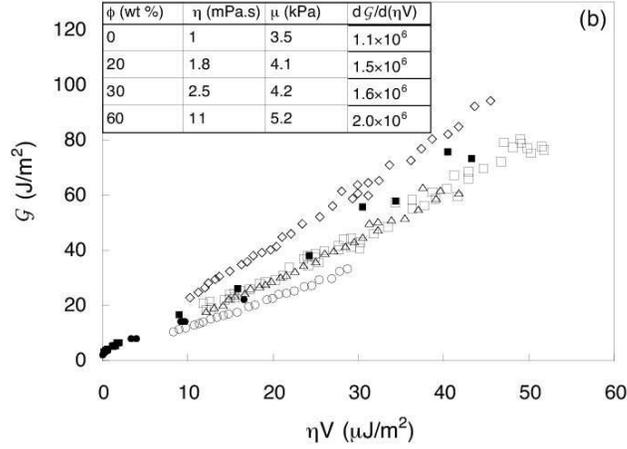}
    \caption{{\bf Influence of solvent viscosity on the fracture 
    energy $\mathcal G(V)$}. {\bf a,} Gels with 
    a constant gelatin content (5 wt \%) in water/glycerol solvents 
    with glycerol concentration $\phi$ = 0 wt \% (circles), 20 wt \% (triangles), 30 
    wt \% 
    (squares), 60 wt \% (diamonds). Filled symbols correspond to 
    stationary cracks,  open symbols to 
    cracks accelerated in response to a steady increase of $\lambda$. 
    $\mathcal G_{0} = 2.5 \pm 0.5$  
    J.m$^{-2}$ is the common linearly extrapolated toughness. 
    {\bf b,} Plot of $\mathcal G$ {\it vs.} $\eta\,V$. Data points have 
    been randomly decimated for clarity. Note the weak systematic growth 
    of $d{\mathcal G}/d(\eta V)$ with glycerol content $\phi$ elastic modulus 
    $\mu$ (see Table).} 
    \label{fig:G(etaV)}
    
\end{figure}

\begin{figure}[tbp]
    \centering
\includegraphics[scale=0.5]{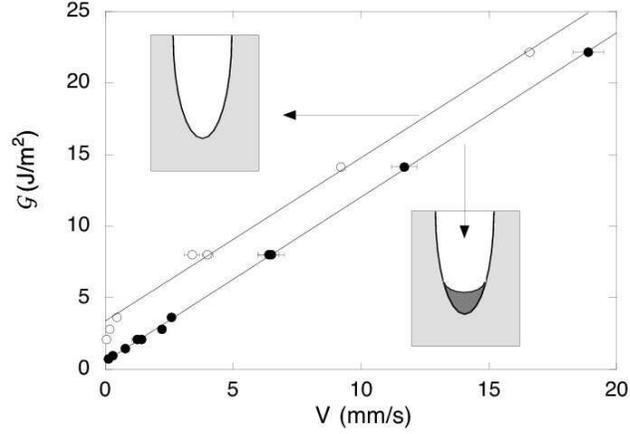}
    \caption{{\bf Effect of tip homo-wetting.} $\mathcal G(V)$ curves for a 5 wt\% gelatin gel in pure 
    water~:  ``dry" 
    cracks opening in 
    ambient air (upper data) and ``wet" 
    cracks with a drop of pure water soaking the tip. Each pair of 
    points for a given 
    $\mathcal G$  corresponds to a 
    single sample.  
Error bars show standard deviations of velocity along the track.
At $\mathcal G$ too low  for dry cracks to 
    propagate, wet ones can still run. Linear fits 
    are shown. The wet data appear merely translated towards lower energies.
    The extrapolated fracture energy for wet tips is $\mathcal 
    G_{0}^{wet} = 0.6 \pm 0.15$ J.m$^{-2}$.
    } 
    \label{fig:Gdry/wet}
\end{figure}

\begin{figure}[tbp]
    \centering
\includegraphics[scale=0.5]{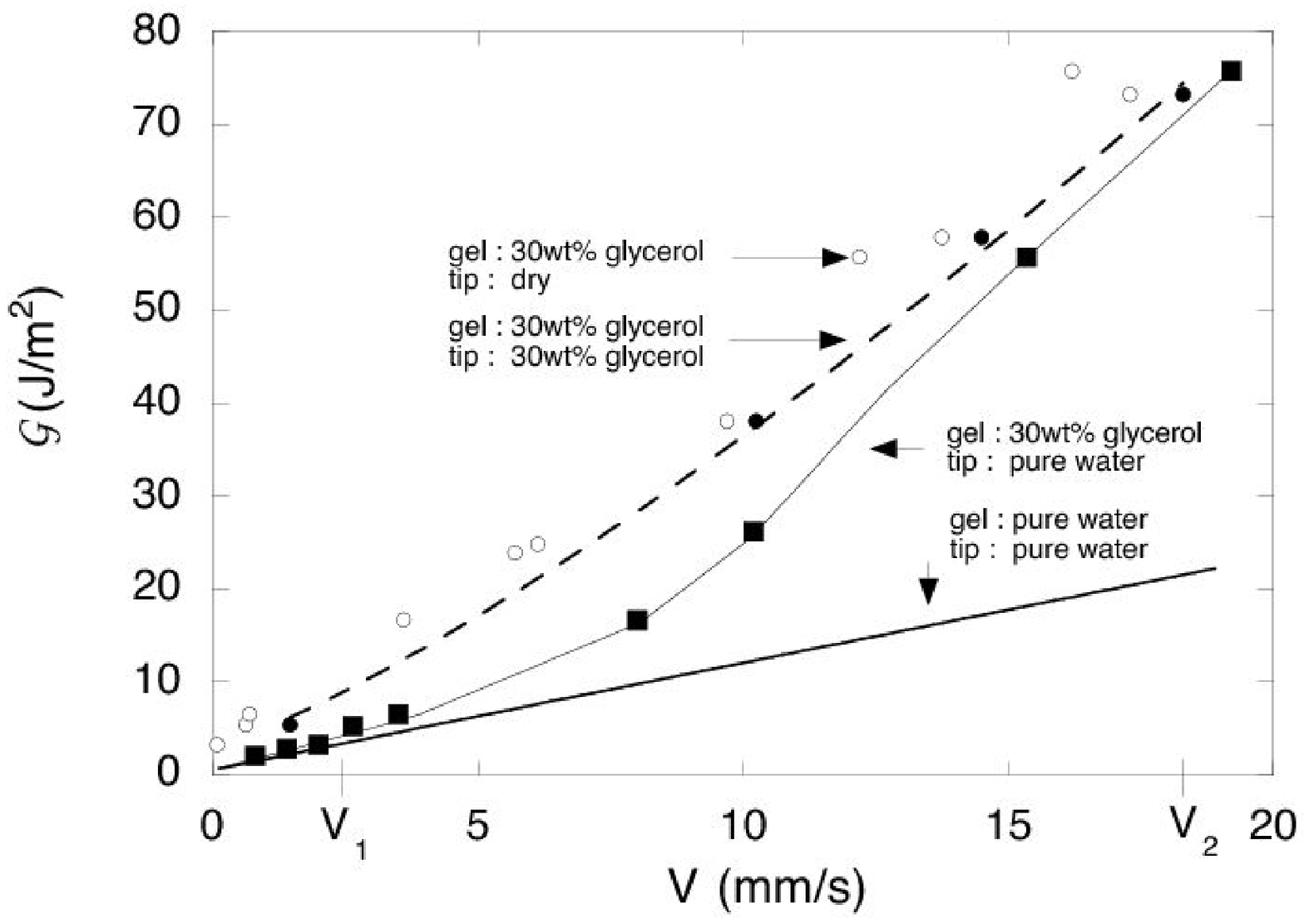}
\caption{{\bf Effect of tip hetero-wetting by a less viscous solvent.} 
The tip of a  crack 
    propagating through a gel of gelatin in $\phi$ = 30\% glycerol/water 
     is wetted with pure water (squares). The full curve is a guide for the eye. 
     At low velocities, the data fall on the 
    curve for a gel of gelatin and {\it pure water} wetted by pure 
    water (full line). At higher 
    velocities ($V_{1}\alt V \alt V_{2}$), the $\mathcal G(V)$ curve crosses over 
    and  approaches 
    the one for the glycerolled 
    gel wetted by the same solvent (closed circles and dash 
    line). 
    The data for a  $\phi$ = 30\% glycerol/water gel fractured in 
    ambient air (open circles) are shown for  
    comparison.}
    \label{fig:HeteroWetting}
\end{figure}

The   energy 
released by unit area of crack advance is computed from the elastic 
energy of the uncracked plate $\mathcal F(\lambda)$ as\cite{Rivlin}~:
$\mathcal G = \mathcal F(\lambda)/(eL)$ whith $e$ = 10 mm  the thickness 
of the plate and $L = 300$ mm its initial length. 
This takes into account elastic non-linearities in the far-field 
region ahead of the crack tip. 
Each plate results in a single $\mathcal G(V)$ data point. 
We have also performed material-saving experiments in which the stretching ratio was 
increased at a constant rate $\dot\lambda = 1.7\times10^{-2}$ s$^{-1}$. 
The resulting non-steady crack velocity along the 
crack path was
computed from a piecewise linear fit. 
We have validated these $\mathcal G(V)$ data by comparison 
with steady-state ones on an overlapping velocity range 
(Fig.~\ref{fig:G(V)}(a)).

Experiments with wetted crack tips are performed by injecting a drop of solvent
of about $250 \mu$l 
 in the tip region while the crack runs. The solvent follows 
the tip and is 
prevented from flowing out both thanks to gravity (the crack running 
down vertically) and capillarity (the solvent wets the gel and forms a 
meniscus bridging the fracture gap).

\newpage

\begin{thebibliography}{}
%
\bibitem{Food} Van Vliet, T. \& Walstra, P. Large deformation and 
fracture behaviour of gels. {\it Faraday Discuss.} {\bf 101,} 
359--370 (1995).
%
\bibitem{Tissue} Lee, K. Y. \& Mooney, D. J. Hydrogels for tissue 
engineering. {\it Chem. Rev.} {\bf101,} 1869--1879 (2001).
%
\bibitem{Mooney} Kong, H. J., Wong, E. \& Mooney, D. J. Independent 
control of rigidity and toughness of polymeric hydrogels. {\it 
Macromolecules} {\bf 36,} 4582--4588 (2003).
%
\bibitem{McEvoy} Mc Evoy, H., Ross-Murphy, S. B. \& Clark, A. H. Large 
deformation and ultimate properties of biopolymer gels: 1. Single 
biopolymer component systems. {\it Polymer} {\bf 26,} 1483--1492 
(1985).
%
\bibitem{Nij} te Nijenhuis, K. Thermoreversible networks. {\it 
Advances in Polymer Science} {\bf 130,} Chap. 10 (1997). 
%
\bibitem{Courty} Courty, S., Gornall, J. L. \& Terentjev, E. M. 
Induced helicity in biopolymer networks under stress. {\it Proc. Nat. 
Acad. Sci} {\bf 102,} 13457--13460 (2005).
%
\bibitem{Onuki} Onuki, A. Theory of phase transition in polymer 
gels. {\it Adv. Polymer Sci.} {\bf 109,} 63--121 (1993).
%
\bibitem{DLJohnson} Johnson, D. L. Elastodynamics of gels. {\it J. Chem. 
Phys} {\bf 77,} 1531--1539 (1982). 
%
\bibitem{Higgs} Higgs, P. G. \& Ross-Murphy, S. B. Creep measurements 
on gelatin gels. {\it Int. J. Biol. Macromol.} {\bf 12,} 233--240 
(1990).
 
%
\bibitem{Bonn} Bonn, D., Kellay, H., Prochnow, M., Ben-Djemiaa, K., \&  
Meunier, J. Delayed Fracture of an Inhomogeneous Soft Solid. {\it 
Science} {\bf 280,}  265--267  (1998).   
%
\bibitem{LT} Lake, G. J. \& Thomas, A.G. The strength of highly elastic 
materials. {\it Proc. R. Soc. London A} {\bf 300,} 108--119 (1967).
%
\bibitem{Nishi} Nishinari, K., Koide, S. \& Ogino, K. On the 
temperature dependence of elasticity of thermo-reversible gels. {\it J. 
Physique} {\bf 46,} 793--797 (1985). 
%
\bibitem{Ball} Higgs, P. G. \& Ball, R. C. Some ideas concerning the 
elasticity of biopolymer networks. {\it Macromolecules} {\bf 22,} 
2432--2437 (1989).
%
\bibitem{Lawn} Lawn, B. R. Fracture of brittle solids --- 2nd edn, 
Cambridge, University Press (1993). 
%
\bibitem{Hui} Hui, C.-Y., Jagota, A., Bennison, S. J. \& Londono, J. D. 
Crack blunting and the strength of soft elastic solids. 
{\it Proc. R. Soc. London A} {\bf 459,} 1489--1516 (2003). 
%
\bibitem{Tanaka} Tanaka, T., Hocker, L.O. \& Benedek, G. B. 
Spectrum of light scattered from a viscoelastic gel. {\it J. Chem. 
Phys.} {\bf 59,} 5151--5159 (1973). 

%
\bibitem{Gelpulses} Baumberger, T., Caroli, C. \& Ronsin, O. Self 
healing pulses and the friction of gelatin gels. {\it Eur. Phys. J. 
E} {\bf 11,} 85--93 (2003).
%
\bibitem{Rivlin} Rivlin, R. S. \& Thomas, A. G. Rupture of rubber. I. 
Characteristic energy for tearing. {\it J. Polymer Sci.} {\bf 10,} 
291--318 (1953). 
%
\bibitem{Ferry} Laurent, J.-L., Janmey, P. A. \& Ferry, J. D. Dynamic 
viscoelastic properties of gelatin gels in glycerol-water mixtures. 
{\it J. Rheol.} {\bf 24,} 87--97 (1980). 

\end{thebibliography}
\end{document}